\documentclass[12pt]{article}
\usepackage[a4paper, margin=1in]{geometry}

\usepackage[english]{babel}
\usepackage{amsmath, amsfonts, amssymb, dsfont, graphicx, tabularx, adjustbox, graphics, bbm, mathrsfs, mathtools, multirow, nameref, natbib, setspace, thmtools, bm, algorithm, algpseudocode}
\usepackage{enumitem}
\usepackage[font=footnotesize,justification=centering]{caption}
\usepackage{subcaption}

\usepackage[normalem]{ulem}

\usepackage{hyperref}

\usepackage{tabularx,booktabs}

\newtheorem{theorem}{Theorem}
\newtheorem{corollary}{Corollary}

\DeclareMathOperator*{\argmin}{argmin}
\DeclareMathOperator{\sign}{sign}

\usepackage{authblk}

\begin{document}
\def\spacingset#1{\renewcommand{\baselinestretch}%
{#1}\small\normalsize} \spacingset{1}

\title{Integrative Learning of Individualized Treatment Rules from Multiple Studies with Partially Overlapping Treatments}

\author{Yuan Bian$^{1}$, Donglin Zeng$^{2}$, Hyun-Joon Yang$^{3}$, Leanne M. Williams$^{3}$, and Yuanjia Wang$^{1,4,*}$\\
$^{1}$Department of Biostatistics, Columbia University\\
$^{2}$Department of Biostatistics, University of Michigan\\
$^{3}$Department of Psychiatry and Behavioral Sciences, Stanford University\\
$^{4}$Department of Psychiatry, Columbia University\\
}
\date{}

\maketitle

\begin{abstract}
An individualized treatment rule (ITR) tailors treatments to a patient’s specific characteristics. However, randomized controlled trials (RCTs) are often underpowered to detect the treatment effect heterogeneity needed for reliable ITR estimation. To address this limitation, there is growing interest in leveraging information from multiple studies to improve statistical power and support individualized decision-making. A key challenge in this context is that available RCTs may not evaluate the same set of treatments. In this paper, we propose an integrative learning framework that synthesizes evidence across multiple RCTs that share a common comparator but differ in their alternative treatment arms. Our method integrates information through a regularized weighted misclassification risk function and adaptively determines the contribution of each study to the ITRs of the others. We rigorously study the excess risk of the resulting estimator. Simulation studies demonstrate that the proposed approaches improve the estimation of both value and benefit functions. We illustrate the utility of our methodology using data from two landmark studies of major depressive disorder: the Establishing Moderators and Biosignatures of Antidepressant Response in Clinical Care study and the International Study to Predict Optimized Treatment in Depression study, both of which include a selective serotonin reuptake inhibitor as a common treatment arm. We find that the separate learning method outperforms one-size-fits-all methods, and our integrative methods further improve performance.\\
\end{abstract}

\noindent%

\doublespacing
\allowdisplaybreaks

\section{Introduction}
Major depressive disorder (MDD) is a chronic, disabling, and often recurrent mood disorder that contributes substantially to global mortality and disease burden \citep{friedrich2017depression, bitter2024mortality}. Despite significant investments in the development of pharmacological treatments, no universally effective therapy for MDD has been identified \citep{saveanu2015international,trivedi2016establishing}. Antidepressant response remains highly heterogeneous and difficult to predict across individuals, suggesting the need for personalized treatment strategies that leverage individual-level characteristics and biomarkers, much like how clinicians personalize treatments for individual patients \citep{liu2016there}.

Randomized controlled trials (RCTs) remain the gold standard for evaluating treatment efficacy and are increasingly used to develop individualized treatment rules (ITRs), which aim to tailor treatments to patient-specific profiles \citep[e.g.,][]{qian2011performance,zhao2012estimating,liu2018augmented}. However, most RCTs are powered to detect average treatment effects and are ill-equipped to identify the treatment effect heterogeneity necessary for reliable ITR estimation. To address this limitation, there is growing interest in integrating data across multiple trials to enhance statistical power and support individualized decision-making.

Due to inter-study heterogeneity, arising from differences in study populations or trial designs \citep{sutton2008recent}, naively applying existing methods to combined studies may lead to biased results and, consequently, misleading scientific conclusions \citep{tang2016fused}. Meta-analysis \citep{whitehead2002meta} is the most widely used approach for combining evidence across studies. Traditional meta-analyses typically rely on summary statistics, which are useful when individual participant data (IPD) are unavailable. However, IPD provides richer information for reliable ITR estimation, and there has been increasing attention to meta-analysis methods that incorporate IPD \citep[e.g.,][]{riley2008meta, debray2013framework}. Such methods, often based on random-effects or multilevel models, can account for between-study variability and allow examining treatment effect heterogeneity, but they often lack interpretability and may suffer from reduced statistical power \citep{tang2016fused}. Integrative data analysis \citep[IDA;][]{curran2009integrative, brown2018two} offers an alternative strategy, pooling IPD from multiple studies and analyzing them as if drawn from a single data-generating process using shared parameter models. 

\citet{bareinboim2016causal} presented a theoretical solution to the problem of data fusion in causal inference tasks. \citet{tang2016fused} proposed a fused Lasso-based approach in the context of data integration, while \citet{shen2020fusion} and \citet{cai2023individualized} considered weighted aggregation of individual information under different asymptotic regimes, demonstrating that incorporating information from similar individuals can improve estimation. \citet{gao2024fusing} and \citet{qiu2024integrative} employed fusing penalties to synthesize evidence across outcomes and trials, respectively, and estimate integrative ITRs. However, these methods typically assume that all studies compare the same set of treatments, an assumption often violated in real-world RCT networks and in the field of mental health.

In the case of MDD, many RCTs compare different pairs of antidepressants. For example, the Establishing Moderators and Biosignatures of Antidepressant Response for Clinical Care (EMBARC) study \citep{trivedi2016establishing} compared sertraline, a selective serotonin reuptake inhibitor (SSRI), with placebo, while the International Study to Predict Optimized Treatment in Depression (iSPOT-D) study \citep{williams2011international} compared sertraline with venlafaxine XR, a serotonin-norepinephrine reuptake inhibitor (SNRI). Although these trials do not share a common treatment contrast, they do share SSRI as a common treatment arm, suggesting that information from one trial could inform decision-making in the other.  For instance, if the iSPOT-D findings indicate that SSRIs are superior to SNRIs in a particular subgroup, the EMBARC ITRs may more confidently favor SSRIs over placebo in that population.  Conversely, if the iSPOT-D ITRs suggest that certain patients respond better to SNRIs than SSRIs, then the ITRs for EMBARC might be more cautious in recommending SSRIs for similar patients. The same logic applies in the opposite direction, where ITRs learned from EMBARC can inform those for iSPOT-D. These observations motivate the need for an integrative framework that accommodates trials with partially overlapping treatments and enables estimation of ITRs informed by evidence from multiple studies.

In this paper, we propose a novel integrative learning framework for synthesizing evidence across multiple RCTs that share a common comparator but differ in their other treatment arms. Our approach jointly estimates study-specific ITRs, and adaptively integrates information across studies via a regularized weighted misclassification risk. We introduce two integrative methods: the first, IntLS, incorporates learned ITRs from the secondary study, while the second, IntLF, fully leverages IPD from both studies within the learning process. We show that the resulting optimization problem can be efficiently solved using the existing outcome-weighted learning algorithms \citep{liu2018augmented}, augmented with supplementary penalty terms that incorporate additional information from the secondary study. Additionally, we examine the excess risks for the resulting estimators. 

Our approach differs from network meta-analysis \citep[NMA;][]{lumley2002network}, which typically focuses on estimating relative treatment effects using contrast- or arm-based models and does not aim to learn patient-level decision rules. It is worth noting that a similar procedure to IntLS was used in \cite{gao2024fusing} and \cite{qiu2024integrative} for integrating ITRs, though the focus differs. \cite{gao2024fusing} proposed a method to integrate multiple outcomes, while \cite{qiu2024integrative} explored the integration of multiple trials with varying resolutions of patient-specific characteristics. In contrast, our work focuses on integrating information with partially overlapping treatments across multiple trials and introduces the enhanced IntLF method with a theoretical analysis of its excess risk. 

The remainder of the paper is organized as follows. Section \ref{sec: m} introduces the separate learning and two proposed integrative learning methods, along with a doubly robust implementation. Section \ref{sec: tr} establishes the theoretical results, followed by simulation studies in Section \ref{sec: ss}. Section \ref{sec: re} illustrates the proposed methods using the EMBARC and iSPOT-D studies. We conclude with a discussion and directions for future research in Section \ref{sec: d}.

\section{Methods}\label{sec: m}
\subsection{Notation and Objective}
Consider two studies that share a common treatment comparator. Study $1$ evaluates treatment $T_1$, which offers two options $A$ (coded as $-1$) and $B$ (coded as $1$), involving $n_1$ subjects with $p$-dimensional tailoring variables $\boldsymbol X_1$ and outcomes $R_1$. Study $2$ compares treatment $T_2$, which involves options $C$ (coded as $-1$) and $B$ (coded as $1$), comprising $n_2$ subjects with $p$-dimensional tailoring variables $\boldsymbol X_2$ and outcomes $R_2$. Treatment $B$ serves as the common comparator across both studies. Without loss of generality, we assume that higher values of $R_j$ indicate more favorable outcomes, reflecting better health status (e.g., improvement of depressive symptoms). 

An {\it individualized treatment rule} (ITR) $d_j:\boldsymbol X_j\to T_j$, where $j=1,2$, is a function that maps a patient's tailoring variables to a recommended treatment. Each ITR can be represented as $d_j(\boldsymbol X_j)=\text{sign}\{{f_j(\boldsymbol X_j)}\}$ for some {\it decision function} $f_j:\boldsymbol X_j\to\mathbb{R}$. In Study $1$, the ITR $d_1$ chooses between treatment options $A$ or $B$; in Study $2$, the ITR $d_2$ selects between $B$ and $C$. Let $R_j(t)$ denote the potential outcome under treatment $t$, and define the value function for ITR $d_j$ as $V_j(d_j)\triangleq \mathbb{E}[R_j\{d_j(\boldsymbol X_j)\}]$, representing the expected outcome when treatments are assigned according to $d_j$. The benefit function for $d_j$ is defined as $V_j(d_j)-V_j(-d_j)$, which quantifies the expected improvement when assigning treatments based on $d_j$, compared to its complementary rule $-d_j$ \citep{qiu2018estimation}. Let $\pi_j(t,\boldsymbol x)=\Pr(T_j=t|\boldsymbol X_j=\boldsymbol x)$ denote the probability of receiving treatment $t$ given tailoring variables $\boldsymbol{x}$ for study $j$ (i.e., {\it propensity score}). In randomized trials, $\pi_j(t,\boldsymbol x)$ is typically known by design, whereas in observational studies it must be estimated and may depend on $\boldsymbol x$. Following \cite{qian2011performance}, the value function $V_j(d_j)$ can be expressed as 
\begin{equation}
\label{eq: v}
\mathbb{E}\left[\frac{R_j\mathbbm{1}\{T_j=d_j(\boldsymbol X_j)\}}{\pi_j(T_j,\boldsymbol X_j)}\right].
\end{equation}
Under standard causal assumptions (including ignorability, consistency, and positivity, see Conditions (C1) - (C3) in Web Appendix A), $V_j(d_j)$ can be estimated from a sample of $n_j$ i.i.d. observations of $\{\boldsymbol X_j,T_j,R_j\}$, denoted by $\{\boldsymbol x_{ij},t_{ij},r_{ij}\}_{i=1}^{n_j}$. Our objective is to integrate ITRs $d_1$ and $d_2$ to improve the estimation of both value and benefit functions in each study, by leveraging shared information across the two studies.

\subsection{Integrative Learning}
The integrated ITR should leverage Study $1$ to assess whether treatment $B$ is better than $A$, and Study $2$ to determine whether $B$ is competitive with $C$, aligning these decisions. In this way, 
%the integrated rules can identify subgroups for which $B$ is optimal among the three options (when both $B>A$ and $B>C$), despite the absence of a study that directly compares all three treatments. For patients where $B$ is not optimal, the recommended treatment may be either $A$ or $C$, although this cannot be directly inferred from either study alone. To address this,
we propose two integrative methods for learning ITRs from multiple studies with partially overlapping treatments. 

Define $Q_j(\boldsymbol x,t)=\mathbb{E}(R_j|\boldsymbol X_j=\boldsymbol x, T_{j}=t)$ and $m_j(\boldsymbol x,d_j)=Q_j(\boldsymbol x,1)\mathbbm{1}\{d_j(\boldsymbol x)=1\}+Q_j(\boldsymbol x,-1)\mathbbm{1}\{d_j(\boldsymbol x)=-1\}$. To enhance statistical efficiency and robustness to model misspecification, we augment the value function \eqref{eq: v} as 
\begin{equation*}
\mathbb{E}\left[\frac{\{R_j-m_j(\boldsymbol X_j, d_j)\}\mathbbm{1}\{T_j=d_j(\boldsymbol X_j)\}}{\pi_j(T_j,\boldsymbol X_j)}+m_j(\boldsymbol X_j, d_j)\right]
\end{equation*}
following \cite{zhang2012robust}. The corresponding estimator is
\begin{equation*}
\text{AIPWE}(d_j)=\frac{1}{n_j}\sum^{n_j}_{i=1}\left[\frac{\left\{r_{ij}-\widehat{m}_j(\boldsymbol x_{ij}, d_j)\}\mathbbm{1}\{t_{ij}=d_j(\boldsymbol x_{ij})\right\}}{\pi_j(t_{ij}, \boldsymbol x_{ij})}+\widehat{m}_j(\boldsymbol x_{ij}, d_j)\right],
\end{equation*}
where $\widehat{m}_j(\boldsymbol x_{ij}, d_j)$ estimates $m_j(\boldsymbol x_{ij}, d_j)$. In observational studies, the propensity score $\pi_j(t_{ij}, \boldsymbol x_{ij})$ must also be estimated. AIPWE$(d_j)$ is a {\it doubly robust} estimator of the value function: it is consistent if either $\pi_j(t_{ij}, \boldsymbol x_{ij})$ or $m_j(\boldsymbol x_{ij}, d_j)$ is correctly specified. In randomized trials, where $\pi_j(t_{ij}, \boldsymbol x_{ij})$ is known, consistency is guaranteed even if $m_j(\boldsymbol x_{ij}, d_j)$ is misspecified.

Define the conditional expectation
\begin{equation*}
g_j(\boldsymbol X_j)\triangleq\mathbb{E}\left\{\left.\frac{\pi_j(-T_j,\boldsymbol X_j)}{\pi_j(T_j,\boldsymbol X_j)}R_j\right|\boldsymbol X_j\right\}=Q_j(\boldsymbol X_j,1)\pi_j(-1,\boldsymbol X_j)+Q_j(\boldsymbol X_j,-1)\pi_j(1,\boldsymbol X_j).
\end{equation*}
Then, note the identity: 
\begin{equation*}
\frac{R_j-m_j(\boldsymbol X_j, d_j)}{\pi_j(T_j,\boldsymbol X_j)}\mathbbm{1}\{T_j=d_j(\boldsymbol X_j)\}+m_j(\boldsymbol X_j, d_j)=\frac{R_j-g_j(\boldsymbol X_j)}{\pi_j(T_j,\boldsymbol X_j)}\mathbbm{1}\{T_j=d_j(\boldsymbol X_j)\}+Q_j(\boldsymbol X_j,-T_j),
\end{equation*}
where the second term on the right-hand side does not involve $d_j$. Hence, maximizing $\text{AIPWE}(d_j)$ is equivalent to minimizing the weighted misclassification risk:
\begin{equation*}
\frac{1}{n_j}\sum^{n_j}_{i=1}\left[\frac{\{r_{ij}-\widehat{g}_j(\boldsymbol x_{ij})\}\mathbbm{1}\{t_{ij}\neq d_j(\boldsymbol x_{ij})\}}{ \pi_j(t_{ij},\boldsymbol x_{ij})}\right],
\end{equation*}
where $\widehat{g}_j(\boldsymbol x_{ij})$ can be obtained from a weighted regression of $r_{ij}$ on $\boldsymbol x_{ij}$, using weights $\pi_j(-t_{ij},\boldsymbol x_{ij})/\pi_j(t_{ij},\boldsymbol x_{ij})$.

Our first integrative method, {\it IntLS} (where ``S'' stands for ``Simple''), builds on this doubly robust estimation framework. For Study $j$, the goal is to minimize the following regularized weighted misclassification risk:
\begin{equation}
    \frac{1}{n_j}\sum^{n_j}_{i=1}\frac{\left\{r_{ij}-\widehat{g}_j(\boldsymbol x_{ij})\right\}\mathbbm{1}\left\{t_{ij}f_j(\boldsymbol x_{ij})<0\right\}}{\pi_j\left(t_{ij},\boldsymbol x_{ij}\right)}+\lambda_j\left\|f_j\right\|^2_\mathcal{H}+\frac{\kappa_j}{n_j}\sum^{n_j}_{i=1}\mathbbm{1}\left\{f_j(\boldsymbol x_{ij})\widehat{f}_{j'}(\boldsymbol x_{ij})<0\right\}, \label{eq: intls}
\end{equation}
where $\mathcal{H}$ denotes a class of candidate decision functions, e.g., a reproducing kernel Hilbert space (RKHS), and $\lambda_j$ and $\kappa_j$ are tuning parameters that can be selected via cross-validation. The second term in \eqref{eq: intls}, involving the RKHS norm $\left\|f_j\right\|_\mathcal{H}$, encourages simpler decision functions through regularization.  For a linear decision function, we have $f_j(\boldsymbol x)=\boldsymbol\alpha_j^\top \boldsymbol x + \beta_j$ and $\left\|f_j\right\|^2_\mathcal{H}=\boldsymbol\alpha_j^\top\boldsymbol\alpha_j$. Alternatively, a nonlinear decision function $f_j$ can be used, such as $f_j(\boldsymbol x)=\sum^{n_j}_{i=1}\alpha_{ji} k_j(\boldsymbol x, \boldsymbol x_{ij}) + \beta_j$, where $k(\cdot,\cdot)$ is a kernel function. A commonly used kernel is the Gaussian radial basis function (RBF) kernel, $k_j(\boldsymbol x, \boldsymbol x')=\exp\left\{-\sigma_j\|\boldsymbol x-\boldsymbol x'\|^2\right\}$. In this case, the norm is given by $\|f_j\|^2_\mathcal{H}=\boldsymbol \alpha_j^\top\boldsymbol K_j \boldsymbol\alpha_j$, where $\boldsymbol\alpha_j=(\alpha_{j1},\ldots,\alpha_{jn_j})^\top$ and $\boldsymbol K_j$ is the $n_j\times n_j$ kernel matrix with $(i, i')$th entry being $k(\boldsymbol x_{ij}, \boldsymbol x_{i'j})$. The tuning parameter $\sigma_j$ is typically chosen as $\sigma_j=1/d_{j,\text{med}}^2$, where $d_{j,\text{med}}$ was the median of the pairwise distances between tailoring variables in Study $j$ \citep[e.g.,][]{wang2018learning}. 

The function $\widehat{f}_{j'}(\cdot)$ in \eqref{eq: intls} represents the decision function learned independently from Study $j'$, using, for example, {\it augmented outcome weighted learning} \citep{liu2018augmented}. We refer to such independently estimated rules as {\it SepL} (separate learning), as they are trained on each study separately. The third term in \eqref{eq: intls} is inspired by the {\it Laplacian penalty} \citep{huang2011sparse} and {\it fusing penalty} \citep{gao2024fusing}, which promotes agreement between ITRs across studies. This penalty leverages the assumption that tailoring variables $\boldsymbol {x} $ have a similar impact on the decision functions across studies, ensuring that the recommended treatment rules are consistent for patients with similar health profiles. For example, suppose for a given patient profile $\boldsymbol x$, a combination therapy is recommended over medication alone in Study $1$ and over psychotherapy alone in Study $2$. These recommendations are consistent in suggesting that combination therapy is preferred for this patient among the three options. Through \eqref{eq: intls}, the decision functions are encouraged to reflect a transitive, logical structure of relative effectiveness of three treatments without directly comparing medication alone with psychotherapy alone.

A key strength of this integrative learning method is its data-adaptive behavior. If information from Study $2$ is useful for improving $V_1(d_1)$ and the assumption between treatment options is met for most patients, the cross-validation procedure will favor a positive value for $\kappa_1$. Conversely, if the auxiliary information is noisy or misleading, the optimal $\kappa_1$ will be near zero, effectively downweighting its influence. The same logic applies symmetrically for $\kappa_2$ and $V_2(d_2)$. This leads to a self-tuning integrative procedure that incorporates or disregards auxiliary information as appropriate, ultimately yielding more reliable and robust ITRs for selecting among the three treatments $A, B, C$, even when no single study compares all three directly.

Following \cite{qiu2024integrative}, the second indicator function in \eqref{eq: intls} can be rewritten as
\begin{equation}
\label{eq: qiu_equivalent}
    \mathbbm{1}\left\{f_j(\boldsymbol x_{ij})\widehat{f}_{j'}(\boldsymbol x_{ij})<0\right\}=\mathbbm{1}\left\{t_{ij}f_j(\boldsymbol x_{ij})<0\right\}\sign\left\{t_{ij}\widehat{f}_{j'}(\boldsymbol x_{ij})\right\}+\mathbbm{1}\left\{t_{ij}\widehat{f}_{j'}(\boldsymbol x_{ij})<0\right\},
\end{equation}
where the last term is independent of $f_j(\cdot)$. Let 
\begin{equation}
\label{eq: por}
\widetilde{r}_{ij}\left(\kappa_j, \widehat{f}_{j'}\right)\triangleq r_{ij}+\kappa_j\pi_j\left(t_{ij},\boldsymbol x_{ij}\right)\sign\left\{t_{ij}\widehat{f}_{j'}(\boldsymbol x_{ij})\right\}
\end{equation} 
and 
\begin{equation}
\label{eq: pod}
\widetilde{\delta}_{ij}\left(\kappa_j, \widehat{f}_{j'}\right)\triangleq \widetilde{r}_{ij}\left(\kappa_j, \widehat{f}_{j'}\right)-\widehat{g}'_j\left(\boldsymbol x_{ij},\kappa_j, \widehat{f}_{j'}\right),
\end{equation} 
where $\widehat{g}'_j$ is estimated analogously to $\widehat{g}_j$, but replacing $r_{ij}$ with $\widetilde{r}_{ij}\left(\kappa_j, \widehat{f}_{j'}\right)$. Then, minimizing \eqref{eq: intls} is equivalent to optimizing the SepL objective function:
\begin{equation}
\label{eq: ints1}
\min_{f_j\in\mathcal{H}}\left[\frac{1}{n_j}\sum^{n_j}_{i=1}\frac{\widetilde{\delta}_{ij}\left(\kappa_j, \widehat{f}_{j'}\right)\mathbbm{1}\left\{t_{ij}f_j(\boldsymbol x_{ij})<0\right\}}{\pi_j\left(t_{ij},\boldsymbol x_{ij}\right)}+\lambda_j\left\|f_j\right\|^2_\mathcal{H}\right],
\end{equation}
enabling the use of the existing algorithm of \cite{liu2018augmented}. Specifically, we flip the signs of both $\widetilde{\delta}_{ij}\left(\kappa_j, \widehat{f}_{j'}\right)$ and $t_{ij}$ when $\widetilde{\delta}_{ij}\left(\kappa_j, \widehat{f}_{j'}\right)<0$, casting \eqref{eq: ints1} to a convex optimization problem. Since minimizing a loss function involving the $0-1$ loss is known to be nondeterministic polynomial-time hard (NP-hard), we replace the $0-1$ loss with a continuous surrogate loss. In our implementation, we use the {\it Huberized hinge loss} \citep{rosset2007piecewise}, defined as $\phi(u)= 0$ if $u\geq1$, $0.25(u-1)^2$ if $-1\leq u<1$, and $-u$ otherwise. This loss is convex, smooth, and free of additional tuning parameters. Thus, the IntLS estimator becomes: 
\begin{equation}
    \widehat{f}_j^\text{\ IntLS}=\argmin_{f_j\in\mathcal{H}} \frac{1}{n_j}\sum^{n_j}_{i=1}\frac{\left|\widetilde{\delta}_{ij}\left(\kappa_j, \widehat{f}_{j'}\right)\right|\phi\left[t_{ij}\sign\left\{\widetilde{\delta}_{ij}\left(\kappa_j, \widehat{f}_{j'}\right)\right\}f_j(\boldsymbol x_{ij})\right]}{\pi\left(t_{ij},\boldsymbol x_{ij}\right)}+\lambda_j\left\|f_j\right\|^2_\mathcal{H}, \label{eq: IntLS-AOL}
\end{equation}
The tuning parameters $\lambda_j$ and $\kappa_j$ can be selected jointly or sequentially, e.g., first tuning $\lambda_j$ from SepL, then fixing it and tuning $\kappa_j$. The procedure is summarized in Algorithm \ref{alg1}.

\begin{algorithm}[h]
\caption{IntLS Algorithm}\label{alg1}
\begin{algorithmic}
\State \textbf{(i)} Estimate $\widehat{g}_j(\boldsymbol{x}_{ij})$ via weighted regression of $r_{ij}$ on $\boldsymbol{x}_{ij}$, using weights $\pi_j(-t_{ij}, \boldsymbol{x}_{ij})/\pi_j(t_{ij}, \boldsymbol{x}_{ij})$;
\State \textbf{(ii)} Compute residuals $\delta_{ij} \triangleq r_{ij} - \widehat{g}_j(\boldsymbol{x}_{ij})$;
\State \textbf{(iii)} Tune $\lambda_j$ to obtain $\widehat{\lambda}_j$, and estimate $\widehat{f}_j^{\ \text{SepL}}$ by solving the following via cross-validation:
\[
\widehat{f}_j^{\ \text{SepL}} = \argmin_{f_j \in \mathcal{H}} \frac{1}{n_j} \sum_{i=1}^{n_j} \frac{\left|\delta_{ij}\right| \phi\left[t_{ij} \, \text{sign}\left\{\delta_{ij}\right\} f_j(\boldsymbol{x}_{ij})\right]}{\pi_j(t_{ij}, \boldsymbol{x}_{ij})} + \lambda_j \|f_j\|_{\mathcal{H}}^2;
\]
\State \textbf{(iv)} Repeat (i), (ii), and (iii) for $j'$ to obtain $\widehat{\lambda}_{j'}$ and $\widehat{f}_{j'}^{\ \text{SepL}}$;
\State \textbf{(v)} Tune $\kappa_j$ to obtain $\widehat{\kappa}_j$, and estimate  $\widehat{f}_j^{\ \text{IntLS}}$ by proceeding as follows via cross-validation:
\begin{itemize}
    \item[] \textbf{(a)} Construct $\widetilde{r}_{ij}\left(\kappa_j, \widehat{f}_{j'}^{\ \text{SepL}}\right)$ as defined in \eqref{eq: por};        
    \item[] \textbf{(b)} Estimate $\widehat{g}'_j\left(\boldsymbol{x}_{ij}, \kappa_j, \widehat{f}_{j'}^{\ \text{SepL}}\right)$ via weighted regression of $\widetilde{r}_{ij}\left(\kappa_j, \widehat{f}_{j'}^{\ \text{SepL}}\right)$ on $\boldsymbol{x}_{ij}$, using the same weights as in step (i);
    \item[] \textbf{(c)} Compute $\widetilde{\delta}_{ij}\left(\kappa_j, \widehat{f}_{j'}^{\ \text{SepL}}\right)$ as defined in \eqref{eq: pod};
    \item[] \textbf{(d)} Tune $\kappa_j$ and estimate  $\widehat{f}_j^{\ \text{IntLS}}$ by solving the following :
    \[
    \widehat{f}_j^{\ \text{IntLS}} = \argmin_{f_j \in \mathcal{H}} \frac{1}{n_j} \sum_{i=1}^{n_j} \frac{\left|\widetilde{\delta}_{ij}\left(\kappa_j, \widehat{f}_{j'}^{\ \text{SepL}}\right)\right| \phi\left[t_{ij} \, \text{sign}\left\{\widetilde{\delta}_{ij}\left(\kappa_j, \widehat{f}_{j'}^{\ \text{SepL}}\right)\right\} f_j(\boldsymbol{x}_{ij})\right]}{\pi_j(t_{ij}, \boldsymbol{x}_{ij})} + \widehat{\lambda}_j \|f_j\|_{\mathcal{H}}^2;
    \]
\end{itemize}
\State \textbf{(vi)} Repeat (v) for $j'$ to obtain $\widehat{\kappa}_{j'}$ and $\widehat{f}_{j'}^{\ \text{IntLS}}$.
\end{algorithmic}
\end{algorithm}

IntLS, however, incorporates only the previously learned ITR, $\widehat{f}_{j'}(\cdot)$, and does not fully leverage available individual-level data from other studies during the learning procedure. To address this limitation, we propose a second method, denoted as {\it IntLF}, which utilizes the available individual-level data from both studies, thus enabling full bidirectional information sharing between studies. Specifically, we aim to minimize the following objective function:
\begin{align*}
    &\frac{1}{n_1}\sum^{n_1}_{i=1}\frac{r_{i1}\mathbbm{1}\left\{t_{i1}f_1(\boldsymbol x_{i1})<0\right\}}{\pi_1\left(t_{i1},\boldsymbol x_{i1}\right)} +\frac{1}{n_2}\sum^{n_2}_{i=1}\frac{r_{i2}\mathbbm{1}\left\{t_{i2}f_2(\boldsymbol x_{i2})<0\right\}}{\pi_2\left(t_{i2},\boldsymbol x_{i2}\right)}+\lambda_{1}\left\|f_1\right\|^2_\mathcal{H}+\lambda_{2}\left\|f_2\right\|^2_\mathcal{H}\\
    &+\frac{\kappa_1}{n_1}\sum^{n_1}_{i=1}\mathbbm{1}\left\{f_1(\boldsymbol x_{i1})f_2(\boldsymbol x_{i1})<0\right\}+\frac{\kappa_2}{n_2}\sum^{n_2}_{i=1}\mathbbm{1}\left\{f_1(\boldsymbol x_{i2})f_2(\boldsymbol x_{i2})<0\right\}.
\end{align*}
To reduce the complexity associated with jointly optimizing all four tuning parameters, we adopt an iterative optimization strategy. Following the rationale used for optimizing IntLS, we decompose the problem and solve the following subproblem separately for each $j$:
\begin{align*}
    \widehat{f}^\text{\ IntLF}_j=\argmin_{f_j\in\mathcal{H}}&\ \frac{1}{n_j}\sum^{n_j}_{i=1}\frac{\left|\widetilde{\delta}_{ij}\left(\kappa_j, \widehat{f}_{j'}\right)\right|\phi\left[t_{ij}\sign\left\{\widetilde{\delta}_{ij}\left(\kappa_j, \widehat{f}_{j'}\right)\right\}f_j(\boldsymbol x_{ij})\right]}{\pi\left(t_{ij},\boldsymbol x_{ij}\right)}\notag\\ &+\lambda_j\left\|f_j\right\|^2_\mathcal{H} +\frac{\kappa_{j'}}{n_{j'}}\sum^{n_{j'}}_{i=1}\phi\left\{f_j(\boldsymbol x_{ij'})\widehat{f}_{j'}(\boldsymbol x_{ij'})\right\},
\end{align*}
where the tuning parameters $\lambda_j$ and $\kappa_j$ can be selected in the same manner as for IntLS. The full algorithmic procedure is summarized in Algorithm \ref{alg2}.

\begin{algorithm}[]
\caption{IntLF Algorithm}\label{alg2}
\begin{algorithmic}
\State \textbf{(i)} Execute Algorithm \ref{alg1};
\State \textbf{(ii)} Tune $\kappa_{j'}$ and estimate $\widehat{f}_j^{\text{\ IntLF}}$ by solving the following via cross-validation:
    \begin{align*}
    \widehat{f}_j^{\text{\ IntLF}} = \argmin_{f_j \in \mathcal{H}} &\frac{1}{n_j} \sum_{i=1}^{n_j} \frac{\left|\widetilde{\delta}_{ij}\left(\widehat{\kappa}_j, \widehat{f}_{j'}^{\text{\ SepL}}\right)\right| \phi\left[t_{ij} \, \text{sign}\left\{\widetilde{\delta}_{ij}\left(\widehat{\kappa}_j, \widehat{f}_{j'}^{\text{\ SepL}}\right)\right\} f_j(\boldsymbol{x}_{ij})\right]}{\pi_j(t_{ij}, \boldsymbol{x}_{ij})} \\
    & + \widehat{\lambda}_j \|f_j\|_{\mathcal{H}}^2 + \frac{\kappa_{j'}}{n_{j'}}\sum^{n_{j'}}_{i=1}\phi\left\{f_j(\boldsymbol x_{ij'})\widehat{f}_{j'}^{\text{\ SepL}}(\boldsymbol x_{ij'})\right\},    
    \end{align*}
    where $\widehat{f}_{j'}^{\text{\ SepL}}(\boldsymbol x_{ij'})$ is estimated in step (iv) of Algorithm \ref{alg1} via cross-validation, $\widetilde{\delta}_{ij}\left(\cdot\right)$ is computed in step (v) of Algorithm \ref{alg1}, $\widehat{\lambda}_j$ is selected in step (iii) of Algorithm \ref{alg1} via cross-validation, and $\widehat{\kappa}_j$ is selected in step (v) of Algorithm \ref{alg1} via cross-validation;
\State \textbf{(iii)} Repeat (ii) to obtain $\widehat{f}_{j'}^{\ \text{IntLF}}$.
\end{algorithmic}
\end{algorithm}

\section{Theoretical Results}\label{sec: tr}
In this section, without loss of generality, we assume that the conditional expectation of $R_j$ has been removed, any negative rewards have had their signs flipped, and the external decision function $f_{j'}$ has been transformed to take the value $1$ if positive and $-1$ if negative. The Bayes optimal ITR is given by $D_j^*(\boldsymbol x_j) = \sign(f_j^*(\boldsymbol x_j))$, where 
$f_j^*(\boldsymbol x_j) = \mathbb{E}\{R_j(1) \mid \boldsymbol X_j = \boldsymbol x_j\} - \mathbb{E}\{R_j(-1) \mid \boldsymbol X_j = \boldsymbol x_j\}$. For a decision function $f$, define the risk function as 
$\mathcal{R}_j(f) \triangleq \mathbb{E}\left[R_j\mathbbm{1}\{T_j f(\boldsymbol X_j)<0\}/\pi_j(T_j,\boldsymbol X_j)\right]$, and augmented $\phi$-risk function as $\mathcal{R}^\phi_j(f) \triangleq \mathbb{E}\left\{\ell^\phi \circ f\right\}$, where  the augmented $\phi$-loss is defined as $\ell^\phi \circ f\triangleq \ell^\phi_1 \circ f + \ell^\phi_2 \circ f + \ell^\phi_3 \circ f$, with $\ell^\phi_1 \circ f \triangleq R_j \phi\{T_j f(\boldsymbol X_j)\}/\pi_j(T_j, \boldsymbol X_j)$, $\ell^\phi_2 \circ f \triangleq \kappa_j \phi\left\{f(\boldsymbol X_j) f_{j'}(\boldsymbol X_j)\right\}$, and $\ell^\phi_3 \circ f \triangleq \kappa_{j'} \phi\left\{f(\boldsymbol X_{j'}) f_{j'}(\boldsymbol X_{j'})\right\}$. We now present the theoretical results for the proposed IntLF estimator using a Gaussian RBF kernel, with the detailed proofs deferred to Web Appendix B.

\begin{theorem}\label{thm1}
Under Conditions (C1) – (C6) in Web Appendix A, and assuming $f_{j'}\in\mathcal{H}$, as $n_j\to\infty$, the excess risk of the estimator $\widehat{f}^\text{\ IntLF}_j$ satisfies 
\begin{equation*}\mathcal{R}_j\left(\widehat{f}^\text{\ IntLF}_j\right) - \mathcal{R}_j(f_j^*) \leq \min\{\mathcal{B}_1, \mathcal{B}_2\},
\end{equation*}
where 
\begin{align*}
\mathcal{B}_1^2=O_p&\left\{\lambda_j^{-1/2}\left(1 + \kappa_j+\kappa_{j'}\right)\sigma_j^{\gamma p/4}n_j^{-1/2} +\lambda_j \sigma_j^p + (1 + \kappa_j + \kappa_{j'})\sigma_j^{-q_{jj'} p}\right\}
\end{align*}
and
\[
\mathcal{B}_2=O_p\left\{\mathcal{R}_j\left(f_{j'}\right) - \mathcal{R}_j\left(f_j^*\right) + \kappa_j^{-1}+n_j^{-1/2}\right\}
\]
with $\gamma > 0$, and $q_{jj'}=\min\left(q_j, q_{j'}\right)$, where $q_j>0$ and $q_{j'}>0$ are defined in Condition (C6).
\end{theorem}

Theorem \ref{thm1} establishes that the excess risk of 
$\widehat{f}^{\ \mathrm{IntLF}}_j$ is governed by the minimum of two bounds. 
The first bound, $\mathcal{B}_1$, reflects the intrinsic 
estimation-approximation trade-off of the RKHS procedure, 
involving the regularization parameter $\lambda_j$, the kernel bandwidth 
$\sigma_j$, and the external borrowing parameters $\kappa_j$ and $\kappa_{j'}$.
The second bound, $\mathcal{B}_2$, captures the potential gain from leveraging 
the external rule $f_{j'}$. We then minimize the upper bound established in Theorem \ref{thm1} to derive the optimal rate.

\begin{corollary}\label{cor1}
Assume the conditions in Theorem \ref{thm1}, the excess risk of $\widehat{f}^\text{\ IntLF}_j$ satisfies
\[
\mathcal{R}_j\!\left(\widehat{f}^{\mathrm{IntLF}}_j\right) 
- \mathcal{R}_j(f_j^*) 
\leq \min\left\{O_p\left((1 + \kappa_j + \kappa_{j'})^{\frac{2 + 4q_{jj'} + \gamma}{2 + 6 q_{jj'} + \gamma}} n_j^{- \frac{q_{jj'}}{2 + 6 q_{jj'} + \gamma}}\right), \mathcal{B}_2\right\}.
\]
Furthermore, suppose the external rule $f_{j'}$ is consistent with the Bayes optimal classifier $f_j^*$ in the sense that $\mathbb{P}\bigl\{ f_{j'}(\boldsymbol X_j) f_j^*(\boldsymbol X_j) < 0 \bigr\} = 0$. If $\kappa_j$ satisfies
$\kappa_j^{-1} = o\bigl(n_j^{-1/2}\bigr)$,
then
\[
\mathcal{R}_j\!\left(\widehat{f}^{\mathrm{IntLF}}_j\right) 
- \mathcal{R}_j(f_j^*) 
\leq O_p\bigl(n_j^{-1/2}\bigr).
\]
\end{corollary}

Corollary \ref{cor1} demonstrates that when the external decision function $f_{j'}$ coincides with the Bayes optimal classifier $f_j^*$ and $\kappa_j$ is chosen sufficiently large so that the term $\kappa_j^{-1}$ is asymptotically negligible, the excess risk of $\widehat{f}^{\ \mathrm{IntLF}}_j$ achieves the rate $n_j^{-1/2}$, improving upon the rate attained by SepL alone, $n_j^{- q_j/(2 + 6 q_j + \gamma)}$. Conversely, if the external decision function provides little useful information (that is, if $f_{j'}$ substantially deviates from $f_j^*$), IntLF reduces to SepL as $\kappa_j$ and $\kappa_{j'}$ approach zero. We expect that the cross-validation procedure in the method will select the optimal values for $\kappa_j$ and $\kappa_{j'}$, adapting to situations where $f_{j'}$ is either close to or far from $f_j^*$. Corollary \ref{cor1} assumes that the external rule is deterministic. However, when it is estimated, we may require $n_{j'}$ to be sufficiently large so that its variability can be neglected.

\section{Simulation Studies}\label{sec: ss}
We assess the finite-sample performance of the proposed methods through simulation studies, considering both linear and nonlinear scenarios. A total of $200$ simulations are conducted for each setting, using a fixed sample size of $n_1=100$ for the first study and varying the sample size of the second study as $n_2\in\{100, 200, 300\}$.

\subsection{Simulation Design and Data Generation}\label{subsec: sddg}
Consider the covariate vector $\boldsymbol X_{ij}=\left(X_{ij1},\ldots,X_{ij10}\right)^\top$, where $X_{ijk}\sim\mathcal N(0, 1)$ for $i=1,\ldots,n_j$, $j = 1, 2$ and $k=1,2,4,\ldots,10$. The third covariate is constructed as $X_{ij3} = 0.8X'_{ij3}+0.2X_{ij1}$, where $X'_{ij3}\sim\mathcal N(0, 1)$, inducing correlation between $X_{ij1}$ and $X_{ij3}$. The treatment assignment for the $i$th subject in the $j$th study, $T_{ij}$, is drawn from a Bernoulli distribution with success probability $0.5$. The outcome for this subject is defined as $R_{ij}=m_j(\boldsymbol X_{ij})+\iota_j(\boldsymbol X_{ij},T_{ij})+\epsilon_i$, where the main effect functions are given by $m_1(\boldsymbol X_{i1})=1+2X_{i11}+X_{i12}^2+X_{i11}X_{i12}$ and $m_2(\boldsymbol X_{i2})=1+2X_{i21}^2+1.5X_{i22}+0.5X_{i21}X_{i22}$. The error terms $\epsilon_i$ follow a standard normal distribution. In the linear interaction scenario, the treatment interaction functions are defined as $\iota_1(\boldsymbol X_{i1},T_{i1})=T_{i1}(0.2-X_{i11}-2X_{i12})$ and $\iota_2(\boldsymbol X_{i2},T_{i2})=T_{i2}(0.2-X_{i21}-2\rho X_{i22})$, with $\rho\in\{0.9,0.7,0.5\}$. In the nonlinear interaction scenario, the treatment effect functions are given by $\iota_1(\boldsymbol X_{i1},T_{i1})=T_{i1}\{-2.2+\exp(X_{i11})+\exp(X_{i12})\}$ and $\iota_2(\boldsymbol X_{i2},T_{i2})=T_{i2}\{-\tau+\exp(X_{i21})+\exp(X_{i22})\}$, where $\tau\in\{2.3,2.4,2.5\}$.

\subsection{Simulation Results}
\begin{figure}
    \centering
    \includegraphics[width=\linewidth]{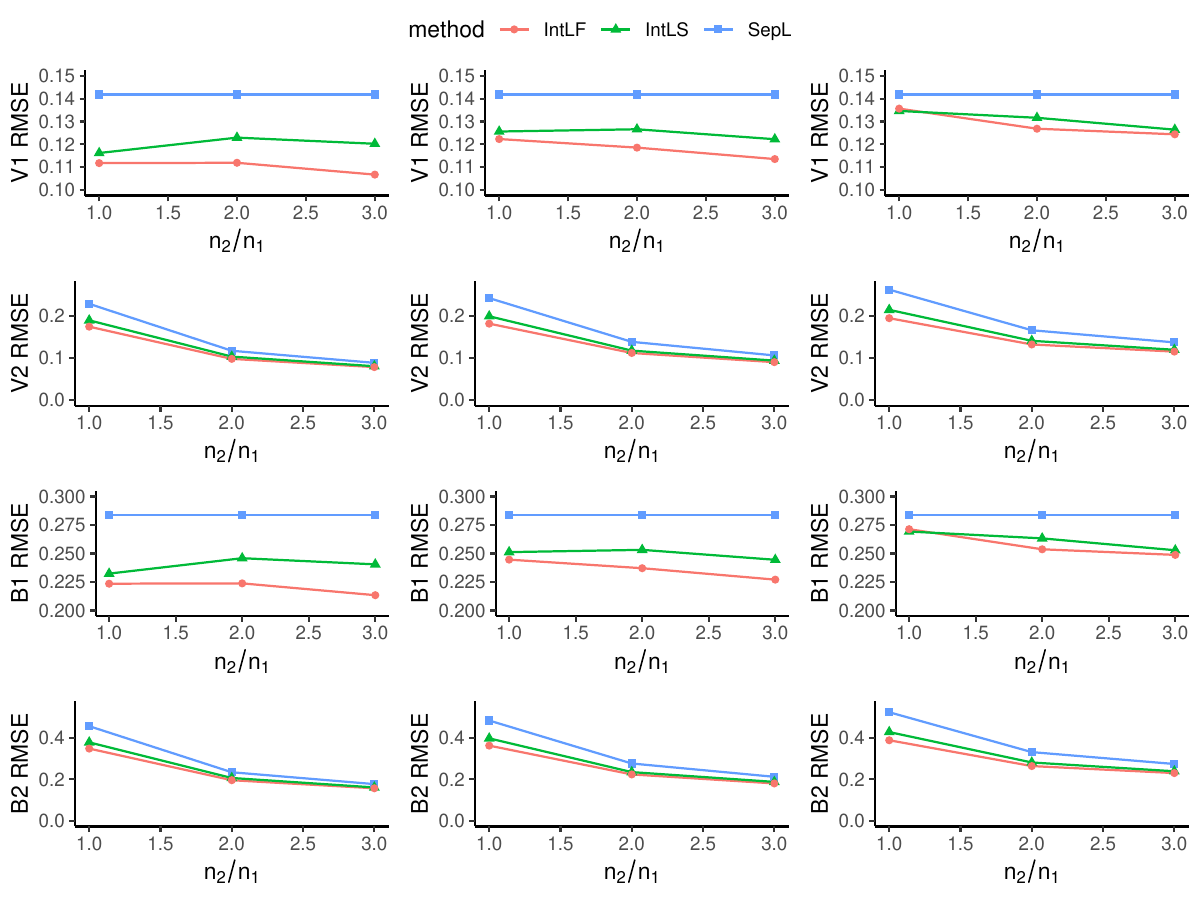}
    \caption{Simulation results for the linear interaction scenario. The first, second, and third columns correspond to $\rho=0.9$, $\rho=0.7$, and $\rho=0.5$, respectively.}
    \label{fig: l}
\end{figure}

\begin{figure}
    \centering
    \includegraphics[width=\linewidth]{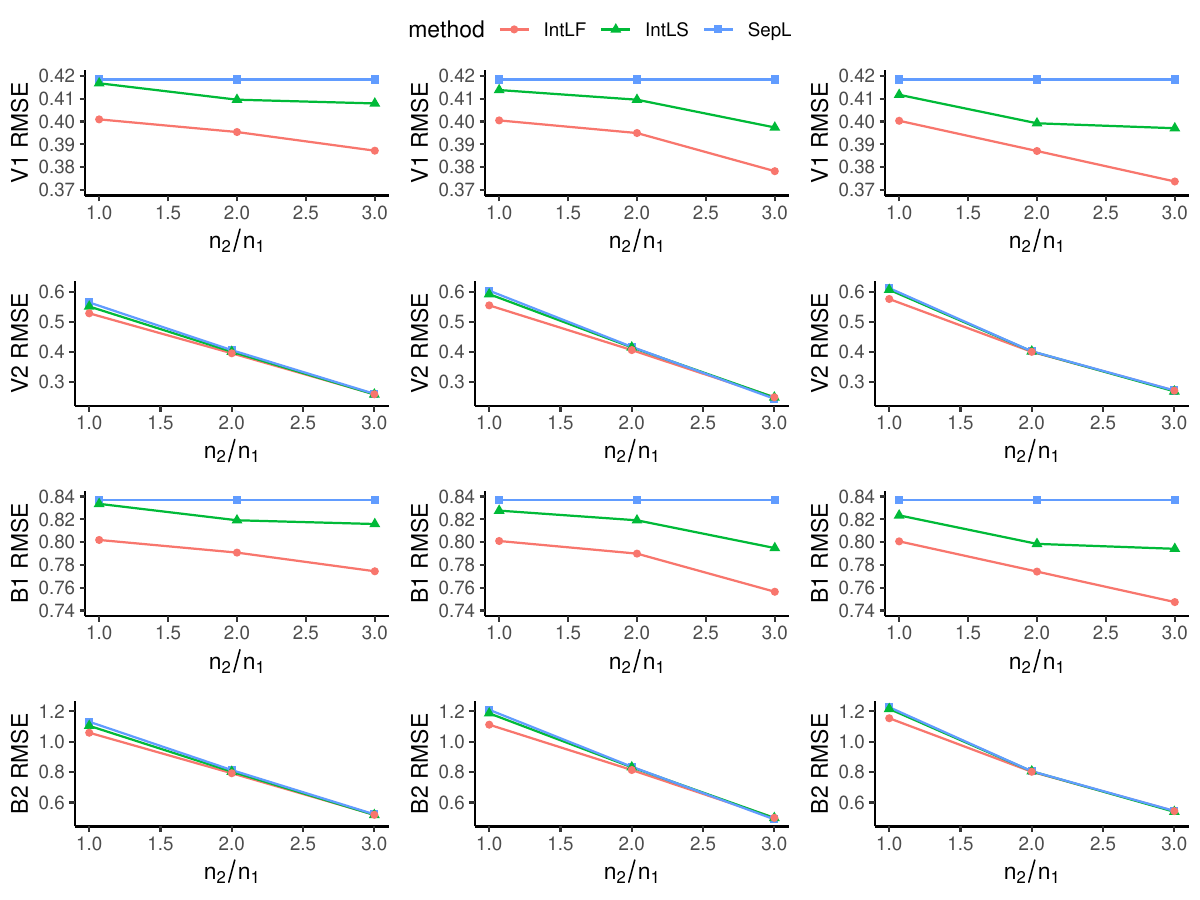}
    \caption{Simulation results for the nonlinear interaction scenario. The first, second, and third columns correspond to $\tau=2.3$, $\tau=2.4$, and $\tau=2.5$, respectively.}
    \label{fig: nl}
\end{figure}

To evaluate the performance of the proposed methods, we compare the estimated value function, denoted $V_j$, and the estimated benefit function, denoted $B_j$, for Study $j$ where $j=1,2$, using a large test set of $100,000$ subjects. The value function quantifies the expected conditional outcome when treatments are assigned according to $d_j(\boldsymbol X_j)$, given a patient’s tailoring variables $\boldsymbol X_j$. In contrast, the benefit function measures the difference in expected conditional outcomes between treatments assigned according to $d_j(\boldsymbol X_j)$ and those assigned according to its complementary rule, $-d_j(\boldsymbol X_j)$, again conditioned on $\boldsymbol X_j$. The root mean square errors (RMSEs) for these two measures, based on $200$ repetitions, are summarized in Figures \ref{fig: l} and \ref{fig: nl}, corresponding to the linear and nonlinear interaction scenarios, respectively. The mean, standard deviation, 2.5\% quantile, and 97.5\% quantile of the bias for these two measures are provided in Web Tables C.1 and C.2, corresponding to the linear and nonlinear interaction scenarios, respectively. Subgroup performance is also presented in Web Figures C.1 and C.2. We select all tuning parameters using 3-fold cross-validation.

For the linear interaction scenario, the parameter $\rho$ represents the similarity between Study $1$ and Study $2$, with $\rho=1$ indicating identical data-generating mechanisms. Smaller values of $\rho$ correspond to increasing divergence between the two studies. In Figure \ref{fig: l}, we examine three levels of similarity: $\rho = 0.9$, $0.7$, and $0.5$. Across these settings, both integrative learning methods, namely IntLS and IntLF, consistently achieve lower RMSEs than SepL on both evaluation metrics, with IntLF typically performing best. However, as the studies become more dissimilar (i.e., as $\rho$ decreases), the performance gains from the integrative methods diminish. For the nonlinear interaction scenario, the similarity parameter is $\tau$, where $\tau=2.2$ denotes identical data-generating mechanisms. Values further from $2.2$ indicate greater divergence. Figure \ref{fig: nl} presents results for $\tau = 2.3$, $2.4$, and $2.5$. Again, IntLS and IntLF generally outperform SepL, with IntLF showing the best performance in most cases.

In both scenarios, SepL's performance in Study $2$ improves with increasing sample size $n_2$. Nevertheless, the integrative methods continue to outperform SepL across all $n_2$ values in the linear scenario. In the nonlinear scenario, integrative methods may perform negligibly worse when $\tau=2.4$ and $n_2=300$. Notably, when the total sample size is held constant (i.e., SepL uses $N$ subjects from Study $2$, and the integrative methods use $N$ subjects split between Studies $1$ and $2$), SepL outperforms IntLS and IntLF on both V2 and B2, for both scenarios. For example, SepL learned on $200$ (or $300$) Study $2$ subjects yields better results on both V2 and B2 than the integrative methods learned on $100$ (or $200$) Study $2$ subjects combined with $100$ subjects from Study $1$.

The subgroup analysis in Web Appendix C indicates that for the two subgroups where the shared treatment is either the best or the worst, the integrative methods outperform the separate method. For the other two subgroups, the integrative methods may perform slightly worse than the separate method, but the difference is minimal. This suggests that the integrative methods may sacrifice some performance in certain subgroups to improve overall performance across the population. This trade-off is expected, as our integrative methods aim to promote consistency in treatment assignments involving the shared comparator.

Finally, for the linear interaction scenario, we also consider a more divergent setting with $\rho = 0.3$. When $n_1=n_2=100$, the integrative methods slightly underperform SepL on Study $1$ but still outperform it on Study $2$. Specifically, the RMSEs for V1 are $0.142$, $0.153$, and $0.162$ for SepL, IntLS, and IntLF, respectively; for B1, the RMSEs are $0.284$, $0.306$, and $0.323$; for V2, $0.283$, $0.243$, and $0.229$; and for B2, $0.566$, $0.486$, and $0.458$. Although cross-validation ideally sets $\kappa_j$ to zero when studies differ substantially, practical constraints, such as small sample sizes, may lead to $\kappa_j>0$, potentially resulting in suboptimal RMSEs.

\section{Real World Applications}\label{sec: re}
We apply our proposed methods to two randomized trials focused on major depressive disorder. The first is the Establishing Moderators and Biosignatures of Antidepressant Response for Clinical Care (EMBARC) study \citep{trivedi2016establishing}, which included $203$ patients randomized to receive either sertraline, a selective serotonin reuptake inhibitor (SSRI), or a placebo. The second is the International Study to Predict Optimized Treatment in Depression (iSPOT-D) study \citep{williams2011international}, which enrolled $337$ patients treated with either sertraline or venlafaxine XR, a serotonin-norepinephrine reuptake inhibitor (SNRI). Both studies share an SSRI as a common treatment arm.

The outcome of interest is the change in total score on the Hamilton Rating Scale for Depression (HRSD) from baseline to post-treatment, with larger positive values indicating greater symptom improvement. The EMBARC study includes 15 HRSD subscores, while the iSPOT-D study includes 17. In EMBARC, the average outcome is $6.817$ for patients on a placebo and $7.670$ for those receiving an SSRI. In iSPOT-D, the averages are $11.459$ for patients given an SNRI and $12.421$ for those on an SSRI. The difference in the averages may be due to EMBARC including two fewer items. To estimate ITRs, we considered three demographic variables: age, gender, and years of education; as well as baseline total scores from the Quick Inventory of Depressive Symptomatology (QIDS) and HRSD. Baseline clinical characteristics and outcomes are summarized in Table \ref{tab: 1}. Electroencephalogram (EEG) features, which have been repeatedly associated with antidepressant response \citep{bruder2013electrophysiological} and previously shown to be informative for learning ITRs in EMBARC \citep{yang2024learning}, were also included. We conducted univariate screening using linear regressions that included each EEG measure, the assigned treatment, and their interaction, selecting the $10$ most informative EEG features. These features represent relative alpha band power across various channels during the eyes-open condition.  

\begin{table}
    \centering
    \caption{Baseline clinical and demographic characteristics and outcomes in the EMBARC and iSPOT-D studies.}
    \label{tab: 1}
    \begin{tabular}{lcccc}
        \toprule
        & \multicolumn{2}{c}{EMBARC ($n_1 = 203$)} & \multicolumn{2}{c}{iSPOT-D ($n_2 = 337$)} \\
        \cmidrule(lr){2-3} \cmidrule(lr){4-5}
        & Mean & SD & Mean & SD \\
        \midrule
        \multicolumn{5}{l}{\textbf{Baseline}} \\
        QIDS & 18.153 & 2.770  & 14.410 & 3.788 \\
        HRSD & 18.443 & 4.316  & 21.935 & 4.154 \\
        Female & 0.640  & 0.481  & 0.588  & 0.493 \\
        Education & 15.196 & 2.448  & 14.401 & 3.478\\
        Age & 37.266 & 13.023 & 37.327 & 12.371 \\
        \addlinespace
        \multicolumn{5}{l}{\textbf{Outcome} (change in HRSD total score)} \\
        Placebo & 6.817 & 7.087 & - & - \\
        SNRI & - & - & 11.459 & 5.815\\
        SSRI & 7.670  & 7.407  & 12.421 & 6.155 \\
        \bottomrule
    \end{tabular}
\end{table}

We randomly split the data into training and test sets in a $1:1$ ratio, used $3$-fold cross-validation to select tuning parameters, and repeated the procedure $100$ times. Our experiments indicate that the linear kernel outperforms the Gaussian RBF kernel, so we present the results using the linear kernel in Table \ref{tab: 2}. We compared the SepL, IntLS, and IntLF methods with baseline one-size-fits-all strategies: All placebo, All SNRI, and All SSRI, which assign the same treatment to all patients. The agreement rates between SepL and both integrative methods are approximately $0.950$. For both studies, the All SSRI strategy yields a better estimated value and benefit compared to All Placebo and All SNRI, respectively. SepL outperforms the All SSRI strategy, and our integrative methods further improve performance. In particular, IntLF achieves higher average values and benefit scores, along with smaller standard errors, compared to IntLS. When using only clinical and demographic features or only EEG features to tailor treatment, the value function and benefit function decrease, suggesting that including both types of features outperforms either type alone.

\begin{table}
    \centering
    \caption{Results for the EMBARC and iSPOT-D studies, based on 100 bootstrap resamples.}
    \label{tab: 2}
    \begin{tabular}{l c c c c c c c}
    \toprule
    \multicolumn{5}{c}{Value and benefit for the outcome (change in HRSD total score)}\\
    \midrule
    & \multicolumn{2}{c}{EMBARC} & \multicolumn{2}{c}{iSPOT-D}\\
    \cmidrule(lr){2-3}\cmidrule(lr){4-5}
    & Value & Benefit & Value & Benefit\\
    \midrule
    All Placebo & 6.833 (0.628) & -0.837 (0.906) & - & -\\
    All SNRI & - & - & 11.485 (0.511) & -0.929 (0.716) \\
    All SSRI & 7.670 (0.805) & 0.837 (0.906)  & 12.414 (0.473) & 0.929 (0.716) \\
    SepL-CD & 7.399 (0.817) & 0.357 (1.156) & 12.329 (0.500) & 0.739 (0.754) \\
    IntLS-CD & 7.359 (0.821) & 0.271 (1.095) & 12.298 (0.507) & 0.673 (0.801)\\
    IntLF-CD & 7.344 (0.845) & 0.245 (1.157) & 12.294 (0.512) & 0.663 (0.795)\\
    SepL-EEG & 7.616 (0.816) & 0.777 (1.246) & 12.316 (0.530) & 0.687 (0.806) \\
    IntLS-EEG & 7.701 (0.822) & 0.961 (1.244) & 12.351 (0.510) & 0.753 (0.732)\\
    IntLF-EEG & 7.718 (0.831) & 0.998 (1.280) & 12.360 (0.517) & 0.771 (0.765)\\
    SepL & 7.726 (0.817) & 1.017 (1.306) & 12.486 (0.513) & 1.035 (0.765) \\
    IntLS & 7.726 (0.871) & 1.011 (1.389) & 12.507 (0.538) & 1.088 (0.779)\\
    IntLF & {\bf 7.733} (0.868) & {\bf 1.023} (1.362) & {\bf 12.515} (0.519) & {\bf 1.101} (0.748)\\
    \bottomrule
    \toprule
    \multicolumn{5}{c}{Agreement rates between IntLs compared to SepL}\\
    \midrule
    & \multicolumn{2}{c}{EMBARC} & \multicolumn{2}{c}{iSPOT-D}\\
    IntLS & \multicolumn{2}{c}{0.958 (0.052)} & \multicolumn{2}{c}{0.954 (0.048)}\\
    IntLF & \multicolumn{2}{c}{0.944 (0.061)} & \multicolumn{2}{c}{0.941 (0.054)} \\
    \bottomrule
    \multicolumn{5}{l}{\footnotesize\textit{Values are shown as mean (SD); SepL-CD, IntLS-CD, and IntLF-CD refer to models that use only clinical }}\\
    \multicolumn{5}{l}{\footnotesize\textit{and  demographic features to tailor treatment. SepL-EEG, IntLS-EEG, and IntLF-EEG refer to models that }}\\
    \multicolumn{5}{l}{\footnotesize\textit{use only EEG features to tailor treatment.}}
    \end{tabular}
\end{table}

When incorporating external outcomes, IntLF also exhibits enhanced clinical relevance. In the EMBARC study, IntLF recommends fewer non-responders, as defined by the Clinical Global Improvement (CGI) scale, compared to SepL and IntLS. The average estimated benefits in response rates defined by CGI between each ITR and its complementary rule are 0.088 for SepL, 0.092 for IntLS, and 0.096 for IntLF, indicating greater discriminatory ability for IntLF. Similarly, in the iSPOT-D study, IntLF yields greater improvements in changes of quality of life, as measured by the World Health Organization Quality of Life (WHOQOL) physical health domain score. The corresponding average estimated benefits for WHOQOL physical health domain scores are 3.687 for SepL, 3.726 for IntLS, and 3.823 for IntLF, again suggesting superior performance of IntLF in identifying clinically beneficial ITRs.

\begin{figure}
    \centering
    \includegraphics[width=\linewidth]{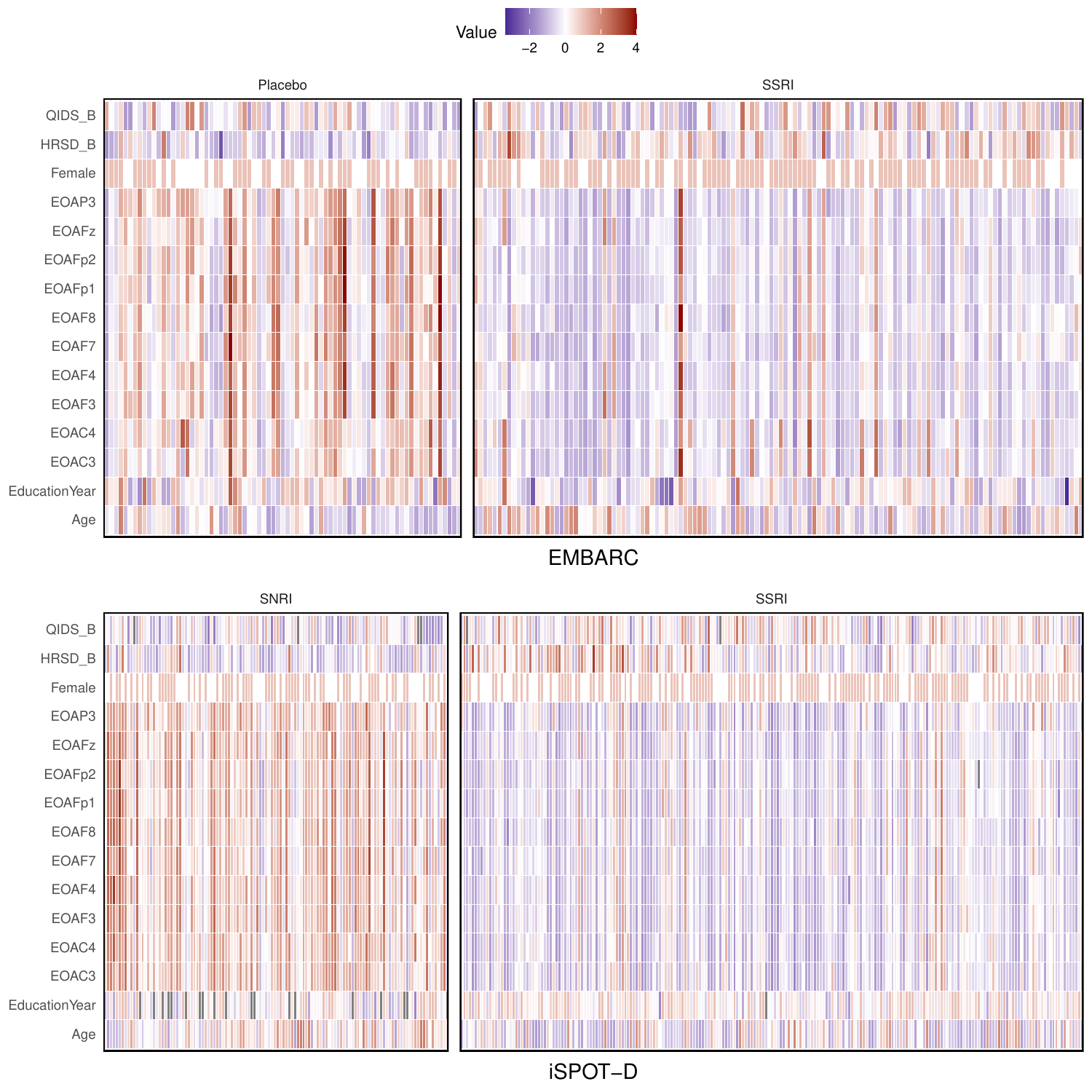}
    \caption{Heatmaps of tailoring variables stratified by the estimated optimal treatment from IntLF, for the EMBARC and iSPOT-D studies. Each column represents a subject, and each row represents a scaled covariate.}
    \label{fig: hm}
\end{figure}

\begin{figure}
    \centering
    \includegraphics[width=\linewidth]{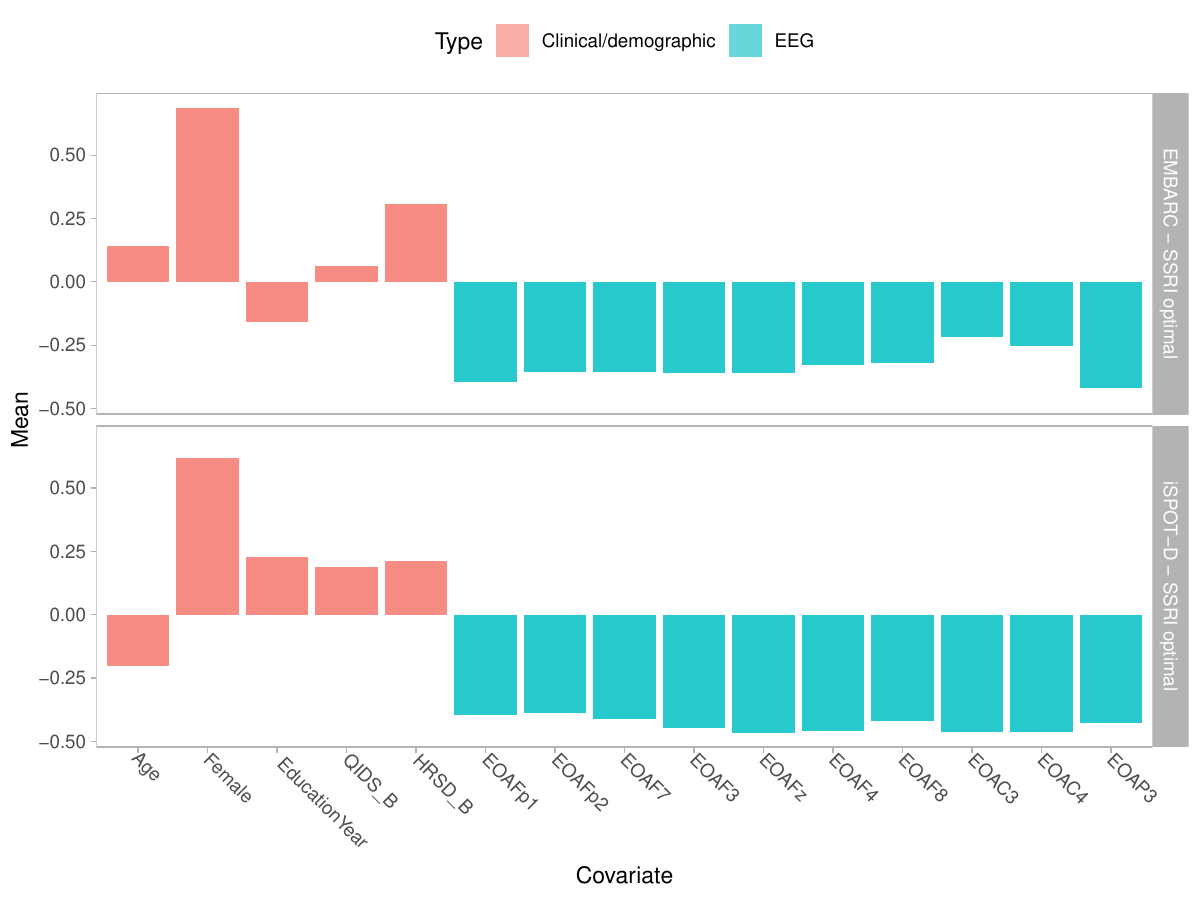}
    \caption{Average scaled tailoring variables for subjects with a predicted optimal treatment of SSRI, as identified by IntLF, in the EMBARC and iSPOT-D studies.}
    \label{fig: barplot}
\end{figure}

Figure \ref{fig: hm} presents heatmaps of the tailoring variables used for learning ITRs across all individuals, and Figure \ref{fig: barplot} displays the average scaled tailoring variables. Both figures are stratified by the predicted optimal treatment from IntLF for each subject in either the EMBARC or iSPOT-D study. The patterns are highly consistent between the two studies. Under IntLF, 63.1\% of patients in EMBARC and 64.4\% in iSPOT-D are recommended to receive an SSRI instead of a placebo or SNRI. This suggests that a substantial subgroup of patients may not benefit from SSRIs, allowing for more targeted treatment and potentially reducing exposure to ineffective therapies. Notably, SSRIs are more likely to be recommended for patients with lower scaled EEG feature values and higher baseline HRSD scores. The association between lower relative alpha band power (as reflected by smaller scaled EEG features) and SSRI recommendation may indicate increased cortical activation or emotional arousal in patients more likely to respond to SSRIs \citep{foxe2011role}. Similarly, higher baseline depression severity, as measured by HRSD, may reflect a clinical profile that benefits more from serotonergic intervention. These patterns highlight the potential value of integrating neurophysiological and clinical markers into treatment decision-making.

\section{Discussion}\label{sec: d}
In this paper, we propose a framework to synthesize evidence across multiple RCTs with a common comparator but differ in their alternative treatment arms. Our proposed method is most effective when the shared comparator is either the best or the worst treatment. When this is not the case, we show that it still outperforms separate learning. However, as studies become more dissimilar (e.g., when many of the optimal treatments are neither the shared treatment nor the alternative), the performance gains from the integrative method may decrease.

Several extensions merit further investigation. One extension is to allow the tuning parameters $\kappa_j$ for the fusion penalties to be subject-specific (i.e., local penalties) as $\kappa_j(\boldsymbol x_{ij})$. For certain subgroups, the consistency constraint may be inapplicable or should be relaxed. When $\kappa_j(\boldsymbol x_{ij})$ is large, disagreement between $f_1$ and $f_2$ is heavily penalized for subjects like $\boldsymbol x_{ij}$. Conversely, when $\kappa_j(\boldsymbol x_{ij})$ is small or zero, the algorithm permits disagreement in that region of the covariate space. To incorporate transfer learning concepts, these subgroups could be identified in a data-driven manner.

The proposed framework was developed for integrating two RCTs to encourage consistent recommendations. In other applications, more than two studies may be fused (e.g., Study $1$: $A$ vs $B$, Study $2$: $B$ vs $C$, and Study 3: $A$ vs $C$). Our methodology naturally extends to such cases by introducing additional fusion penalties for each pair of ITRs expected to be consistent. An extension along the lines of \cite{shen2026two} would also be of interest. For the real data analysis, we consider the change in HRSD scores as the outcome. It is also interesting to explore remission or response as outcomes and borrow information from these measures \citep{gao2024fusing}. These outcomes are binary, and further research is needed to study them more carefully.

Our current framework does not explicitly model or correct for distribution shift; the proposed methods, however, partially mitigate these risks by borrowing information adaptively across studies rather than enforcing full pooling. \cite{chen2024robust} and \cite{chen2024optimizing} studied ITR learning under covariate shift without posterior shift. An extension in this direction using weighting would be valuable. Additionally, it would be interesting to extend our framework to better accommodate algorithmic fairness, specifically to improve ITR performance across all subgroups.

To enhance performance in the presence of irrelevant covariates or high-dimensional covariate spaces, variable selection techniques may be employed. For linear decision functions, the $l_2$ penalty can be replaced with an $l_1$ \citep{tibshirani1996regression} or elastic net \citep{zou2005regularization} penalty. For nonlinear decision functions, scaling covariates and applying an $l_1$ penalty to the scaling parameters can be used to perform implicit variable selection \citep{zhou2017residual}.

\section*{Acknowledgments}
The authors thank the reviewer, the associate editor, and the co-editor for their valuable comments. This manuscript reflects the views of the authors and may not reflect the opinions or views of the National Institutes of Health (NIH) or of the Submitters submitting original data to National Institute of Mental Health Data Archive. The EMBARC data can be obtained through an application at \url{https://dx.doi.org/10.15154/e99n-np66}. The iSPOT-D data (ClinicalTrials.gov identifier: NCT00693849) was sponsored by Brain Resource Company Operations Pty Ltd.

\section*{Supplementary Materials}
Supplementary material is available at Biometrics online.

Web Appendices, Tables, and Figures referenced in Sections \ref{sec: m} - \ref{sec: ss} and code are available with this paper at the Biometrics website on Oxford Academic.

\section*{Funding}
This research is supported by U.S. NIH grants NS073671, GM124104, and MH123487.

\bibliographystyle{apalike}
\bibliography{reference}

@article{tibshirani1996regression,
  title={Regression shrinkage and selection via the lasso},
  author={Tibshirani, Robert},
  journal={Journal of the Royal Statistical Society Series B: Statistical Methodology},
  volume={58},
  number={1},
  pages={267--288},
  year={1996},
  publisher={Oxford University Press}
}

@article{lumley2002network,
  title={Network meta-analysis for indirect treatment comparisons},
  author={Lumley, Thomas},
  journal={Statistics in Medicine},
  volume={21},
  number={16},
  pages={2313--2324},
  year={2002},
  publisher={Wiley Online Library}
}

@book{whitehead2002meta,
  title={Meta-analysis of Controlled Clinical Trials},
  author={Whitehead, Anne},
  year={2002},
  publisher={Wiley},
  address={West Sussex}
}

@article{zou2005regularization,
  title={Regularization and variable selection via the elastic net},
  author={Zou, Hui and Hastie, Trevor},
  journal={Journal of the Royal Statistical Society Series B: Statistical Methodology},
  volume={67},
  number={2},
  pages={301--320},
  year={2005},
  publisher={Oxford University Press}
}

@article{rosset2007piecewise,
  title={Piecewise linear regularized solution paths},
  author={Rosset, Saharon and Zhu, Ji},
  journal={The Annals of Statistics},
  volume={35},
  number={3},
  pages={1012--1030},
  year={2007},
  publisher={JSTOR}
}

@article{riley2008meta,
  title={Meta-analysis of continuous outcomes combining individual patient data and aggregate data},
  author={Riley, Richard D and Lambert, Paul C and Staessen, Jan A and Wang, Jiguang and Gueyffier, Francois and Thijs, Lutgarde and Boutitie, Florent},
  journal={Statistics in Medicine},
  volume={27},
  number={11},
  pages={1870--1893},
  year={2008},
  publisher={Wiley Online Library}
}

@article{sutton2008recent,
  title={Recent developments in meta-analysis},
  author={Sutton, Alexander J and Higgins, Julian PT},
  journal={Statistics in Medicine},
  volume={27},
  number={5},
  pages={625--650},
  year={2008},
  publisher={Wiley Online Library}
}

@article{curran2009integrative,
  title={Integrative data analysis: The simultaneous analysis of multiple data sets},
  author={Curran, Patrick J and Hussong, Andrea M},
  journal={Psychological Methods},
  volume={14},
  number={2},
  pages={81--100},
  year={2009},
  publisher={American Psychological Association}
}

@article{foxe2011role,
  title={The role of alpha-band brain oscillations as a sensory suppression mechanism during selective attention},
  author={Foxe, John J and Snyder, Adam C},
  journal={Frontiers in Psychology},
  volume={2},
  year={2011},
  note={Article 154},
  publisher={Frontiers Research Foundation}
}

@article{huang2011sparse,
  title={The sparse {L}aplacian shrinkage estimator for high-dimensional regression},
  author={Huang, Jian and Ma, Shuangge and Li, Hongzhe and Zhang, Cun-Hui},
  journal={The Annals of Statistics},
  volume={39},
  number={4},
  pages={2021--2046},
  year={2011}
}

@article{qian2011performance,
  title={Performance guarantees for individualized treatment rules},
  author={Qian, Min and Murphy, Susan A},
  journal={The Annals of Statistics},
  volume={39},
  number={2},
  pages={1180--1210},
  year={2011}
}

@article{williams2011international,
  title={International Study to Predict Optimized Treatment for Depression ({iSPOT-D}), a randomized clinical trial: Rationale and protocol},
  author={Williams, Leanne M and Rush, A John and Koslow, Stephen H and Wisniewski, Stephen R and Cooper, Nicholas J and Nemeroff, Charles B and Schatzberg, Alan F and Gordon, Evian},
  journal={Trials},
  volume={12},
  pages={1--17},
  year={2011},
  publisher={Springer}
}

@article{zhao2012estimating,
  title={Estimating individualized treatment rules using outcome weighted learning},
  author={Zhao, Yingqi and Zeng, Donglin and Rush, A John and Kosorok, Michael R},
  journal={Journal of the American Statistical Association},
  volume={107},
  number={499},
  pages={1106--1118},
  year={2012},
  publisher={Taylor \& Francis}
}

@article{zhang2012robust,
  title={A robust method for estimating optimal treatment regimes},
  author={Zhang, Baqun and Tsiatis, Anastasios A and Laber, Eric B and Davidian, Marie},
  journal={Biometrics},
  volume={68},
  number={4},
  pages={1010--1018},
  year={2012},
  publisher={Oxford University Press}
}

@incollection{bruder2013electrophysiological,
    author = {Bruder, Gerard E and Tenke, Craig E and Kayser, J{\"u}rgen},
    title = {Electrophysiological predictors of clinical response to antidepressants},
    booktitle = {The Clinical Handbook for the Management of Mood Disorders},
    publisher = {Cambridge University Press},
    editor = {Mann, J John and McGrath, Patrick J and Roose, Steven P},
    year = 2013,
address = {New York}
}

@article{debray2013framework,
  title={A framework for developing, implementing, and evaluating clinical prediction models in an individual participant data meta-analysis},
  author={Debray, Thomas PA and Moons, Karel GM and Ahmed, Ikhlaaq and Koffijberg, Hendrik and Riley, Richard David},
  journal={Statistics in Medicine},
  volume={32},
  number={18},
  pages={3158--3180},
  year={2013},
  publisher={Wiley Online Library}
}

@article{saveanu2015international,
  title={The international Study to Predict Optimized Treatment in Depression (i{SPOT}-{D}): Outcomes from the acute phase of antidepressant treatment},
  author={Saveanu, Radu and Etkin, Amit and Duchemin, Anne-Marie and Goldstein-Piekarski, Andrea and Gyurak, Anett and Debattista, Charles and Schatzberg, Alan F and Sood, Satish and Day, Claire VA and Palmer, Donna M and others},
  journal={Journal of Psychiatric Research},
  volume={61},
  pages={1--12},
  year={2015},
  publisher={Elsevier}
}

@article{bareinboim2016causal,
  title={Causal inference and the data-fusion problem},
  author={Bareinboim, Elias and Pearl, Judea},
  journal={Proceedings of the National Academy of Sciences},
  volume={113},
  number={27},
  pages={7345--7352},
  year={2016},
  publisher={National Academy of Sciences}
}

@article{liu2016there,
  title={There is individualized treatment. {W}hy not individualized inference?},
  author={Liu, Keli and Meng, Xiao-Li},
  journal={Annual Review of Statistics and Its Application},
  volume={3},
  number={1},
  pages={79--111},
  year={2016},
  publisher={Annual Reviews}
}

@article{tang2016fused,
  title={Fused Lasso Approach in Regression Coefficients Clustering--Learning Parameter Heterogeneity in Data Integration},
  author={Tang, Lu and Song, Peter XK},
  journal={Journal of Machine Learning Research},
  volume={17},
  number={113},
  pages={1--23},
  year={2016}
}

@article{trivedi2016establishing,
  title={Establishing moderators and biosignatures of antidepressant response in clinical care ({EMBARC}): Rationale and design},
  author={Trivedi, Madhukar H and McGrath, Patrick J and Fava, Maurizio and Parsey, Ramin V and Kurian, Benji T and Phillips, Mary L and Oquendo, Maria A and Bruder, Gerard and Pizzagalli, Diego and Toups, Marisa and others},
  journal={Journal of Psychiatric Research},
  volume={78},
  pages={11--23},
  year={2016},
  publisher={Elsevier}
}

@article{friedrich2017depression,
  title={Depression is the leading cause of disability around the world},
  author={Friedrich, Mary Jane},
  journal={JAMA},
  volume={317},
  number={15},
  pages={1517},
  year={2017},
  publisher={American Medical Association}
}

@article{zhou2017residual,
  title={Residual weighted learning for estimating individualized treatment rules},
  author={Zhou, Xin and Mayer-Hamblett, Nicole and Khan, Umer and Kosorok, Michael R},
  journal={Journal of the American Statistical Association},
  volume={112},
  number={517},
  pages={169--187},
  year={2017},
  publisher={Taylor \& Francis}
}

@article{brown2018two,
  title={Two-year impact of prevention programs on adolescent depression: An integrative data analysis approach},
  author={Brown, C Hendricks and Brincks, Ahnalee and Huang, Shi and Perrino, Tatiana and Cruden, Gracelyn and Pantin, Hilda and Howe, George and Young, Jami F and Beardslee, William and Montag, Samantha and others},
  journal={Prevention Science},
  volume={19},
  pages={74--94},
  year={2018},
  publisher={Springer}
}

@article{liu2018augmented,
  title={Augmented outcome-weighted learning for estimating optimal dynamic treatment regimens},
  author={Liu, Ying and Wang, Yuanjia and Kosorok, Michael R and Zhao, Yingqi and Zeng, Donglin},
  journal={Statistics in Medicine},
  volume={37},
  number={26},
  pages={3776--3788},
  year={2018},
  publisher={Wiley Online Library}
}

@article{qiu2018estimation,
  title={Estimation and evaluation of linear individualized treatment rules to guarantee performance},
  author={Qiu, Xin and Zeng, Donglin and Wang, Yuanjia},
  journal={Biometrics},
  volume={74},
  number={2},
  pages={517--528},
  year={2018},
  publisher={Oxford University Press}
}

@article{wang2018learning,
  title={Learning optimal personalized treatment rules in consideration of benefit and risk: With an application to treating type 2 diabetes patients with insulin therapies},
  author={Wang, Yuanjia and Fu, Haoda and Zeng, Donglin},
  journal={Journal of the American Statistical Association},
  volume={113},
  number={521},
  pages={1--13},
  year={2018},
  publisher={Taylor \& Francis}
}

@article{shen2020fusion,
  title={i{F}usion: Individualized fusion learning},
  author={Shen, Jieli and Liu, Regina Y and Xie, Min-ge},
  journal={Journal of the American Statistical Association},
  volume={115},
  number={531},
  pages={1251--1267},
  year={2020},
  publisher={Taylor \& Francis}
}

@article{cai2023individualized,
  title={Individualized group learning},
  author={Cai, Chencheng and Chen, Rong and Xie, Min-ge},
  journal={Journal of the American Statistical Association},
  volume={118},
  number={541},
  pages={622--638},
  year={2023},
  publisher={Taylor \& Francis}
}

@article{bitter2024mortality,
  title={Mortality in patients with major depressive disorder: A nationwide population-based cohort study with 11-year follow-up},
  author={Bitter, Istvan and Szekeres, Gyorgy and Cai, Qian and Feher, Laszlo and Gimesi-Orszagh, Judit and Kunovszki, Peter and El Khoury, Antoine C and Dome, Peter and Rihmer, Zoltan},
  journal={European Psychiatry},
  volume={67},
  number={1},
  pages={e63},
  year={2024},
  publisher={Cambridge University Press}
}

@article{chen2024optimizing,
  title={Optimizing personalized treatments for targeted patient populations across multiple domains},
  author={Chen, Yuan and Zeng, Donglin and Wang, Yuanjia},
  journal={The International Journal of Biostatistics},
  volume={20},
  number={2},
  pages={437--453},
  year={2024},
  publisher={De Gruyter}
}

@article{chen2024robust,
  title={Robust sample weighting to facilitate individualized treatment rule learning for a target population},
  author={Chen, Rui and Huling, Jared D and Chen, Guanhua and Yu, Menggang},
  journal={Biometrika},
  volume={111},
  number={1},
  pages={309--329},
  year={2024},
  publisher={Oxford University Press}
}

@incollection{qiu2024integrative,
  title={Integrative Learning to Combine Individualized Treatment Rules from Multiple Randomized Trials},
  author={Qiu, Xin and Zeng, Donglin and Wang, Yuanjia},
  booktitle={Statistics in Precision Health: Theory, Methods and Applications},
  year={2024},
  editor={Yichuan Zhao and Ding-Geng Chen},
  publisher={Springer}
}

@inproceedings{gao2024fusing,
  title={Fusing individualized treatment rules using secondary outcomes},
  author={Gao, Daiqi and Wang, Yuanjia and Zeng, Donglin},
  booktitle={The 27th International Conference on Artificial Intelligence and Statistics},
  year={2024}
}

@article{yang2024learning,
  title={Learning optimal biomarker-guided treatment policy for chronic disorders},
  author={Yang, Bin and Guo, Xingche and Loh, Ji Meng and Wang, Qinxia and Wang, Yuanjia},
  journal={Statistics in Medicine},
  volume={43},
  number={14},
  pages={2765--2782},
  year={2024},
  publisher={Wiley Online Library}
}

@article{shen2026two,
  title={Two-stage Bayesian network meta-analysis of individualized treatment rules for multiple treatments with siloed data},
  author={Shen, Junwei and Moodie, Erica EM and Golchi, Shirin},
  journal={Statistical Methods in Medical Research},
  number={1},
  pages={3--20},
  volume={35},
  year={2026}
}

\end{document}